# Quasiperiodic Multicolor Solitons in Quasi-Phase-Matched Quadratic Media

## Zhen Qi

*Department of Computer Science and Electrical Engineering,*

*University of Maryland at Baltimore County, Baltimore, MD 21250, USA*

**Abstract**

We study the (1+1)-dimensional quasiperiodic multicolor solitons due to cascading quadratic nonlinear response in generalized one-dimensional quasiperiodic optical superlattice waveguides and show that the dynamic equations describing the quasi-phase-matched multicolor solitons include quasiperiodicity-induced Kerr effects, such as self- and cross-phase modulation, third harmonic generation and four-wave mixing. We demonstrate the stability of this multicolor solitons by means of a Lyapunov analysis based on the energy integral of the wave coupling equations and investigate the dynamics of the multicolor solitons using a virial identity, which predicts a stable propagation of the mutually trapped solitons. We finally establish the analytic stability criterion for the multicolor solitons by applying a multiscale asymptotic method.

## I. INTRODUCTION

Solitons (or solitary waves) play an important role in the dynamics of dispersive wave systems where dispersion (or diffraction) is balanced by nonlinearity. The intense study of soliton effects in nonlinear optics has offered new facilities for all-optical signal processing as well as long-distance communication systems. One of the typical effects is the self-focusing (or self-defocusing) effects and soliton-like beam or pulse propagation produced by the third-order nonlinear response of the so-called $\chi^{(3)}$ materials. Meanwhile, as has been already established, solitary waves can also exist in media with quadratic (or $\chi^{(2)}$) nonlinearity as a result of cascading, under the condition of phase matching [1-7]. Due to the possibility to employ large second-order nonlinearities, such solitary waves, the so-called *quadratic solitons*, have attracted growing attention for the needs of all-optical switching. However, low efficiency of the cascaded nonlinearities was obtained experimentally as a result of the limitations imposed by the use of conventional phase-matching techniques based on birefringence and temperature tuning.

In the context of quadratic nonlinear processes, by introducing at least one additional grating wave vector to compensate for the mismatch between the wave vectors of different waves, the quasi-phase-matching (QPM) technique is known as an attractive way to obtain good phase matching [8-10]. Based on periodic modulations of nonlinear coefficient, the QPM solitons phase-matched by grating wave vectors have been investigated intensively [11-15]. Furthermore, the quasiperiodic solitons, which describes spatially localized self-trapping of a quasiperiodic wave, were generated in a Fibonacci-based quasiperiodic configuration [16]. The multistep phase-matched interaction, for example the third harmonic generation (THG) through cascaded second harmonic generation (SHG) and sum-frequency generation (SFG) processes, can be phase matched simultaneously by multiple grating wave vectors provided by the aperiodic modulation of the nonlinear coefficient, such as a generalized quasiperiodic configuration [17,18]. These parametric interactions lead to simultaneous trapping of all the interacting waves and formation of a multicolor soliton [19-21]. Lobanov and Sukhorukov developed a theoretical description of the multicolor soliton in a one-dimensional (1D) periodic QPM structure, where the two nonlinear processes were realized at quasi-phase-matching of the first order and a higher order [22,23].

In this paper, we present a theoretical analysis of the multicolor soliton formation and stability upon propagation in a quasiperiodically-poled planar QPM waveguide grating, by taking into account the higher-order nonlinearities induced by the quasiperiodicity of the grating. The paper is organized as follows. In the following section we introduce the mathematical model that describes the nonlinear interaction between three cw beams propagating in a generalized quasiperiodic QPM waveguide grating. Then, in Sec. III, we find, numerically, the multicolor solitons that are stable upon propagation in the quasiperiodic QPM waveguide grating and the influence of the induced third-order nonlinearities on their properties. Furthermore, we investigate the excitation of these multicolor solitons, using Gaussian beams. In Sec. IV, a detailed stability analysis of the multicolor solitons based on a Lyapunov method is presented. We investigate the dynamics of the multicolor solitons using a virial identity, which predicts a stable propagation of the multicolor solitons. The stability criterion for the multicolor solitons is established by applying a multiscale asymptotic method. Finally, we conclude with a summary and discussion of our results.

## II. MATHEMATICAL MODEL

We consider the interaction of a cw beam with the fundamental frequency (FF), its second harmonic (SH) and its third harmonic (TH) propagating in a quasiperiodic QPM slab waveguide, where only the quadratic nonlinear susceptibility is modulated. Assuming zero absorption of all interaction waves and introducing the effect of diffraction, the evolution of slowly varying beam envelope is governed by the dynamical equations

$$ik_1 \frac{\partial A_1}{\partial z} + \frac{1}{2}\frac{\partial^2 A_1}{\partial x^2} + k_{10}^2 \chi^{(2)} d(z)\left(A_3 A_2^* e^{-i\Delta_3 z} + A_2 A_1^* e^{-i\Delta_2 z}\right) = 0, \quad (1)$$

$$ik_2 \frac{\partial A_2}{\partial z} + \frac{1}{2}\frac{\partial^2 A_2}{\partial x^2} + k_{20}^2 \chi^{(2)} d(z)\left(A_3 A_1^* e^{-i\Delta_3 z} + \frac{1}{2} A_1^2 e^{i\Delta_2 z}\right) = 0, \quad (2)$$

$$ik_3 \frac{\partial A_3}{\partial z} + \frac{1}{2}\frac{\partial^2 A_3}{\partial x^2} + k_{30}^2 \chi^{(2)} d(z) A_2 A_1 e^{i\Delta_3 z} = 0, \quad (3)$$

where $A_1$, $A_2$ and $A_3$ are the complex electric-field envelopes of the fundamental, the second, and the third harmonic, respectively; $k_1$, $k_2$, and $k_3$ are the wave-vectors of the fundamental, the second, and the third harmonic, respectively; $k_{10}$, $k_{20}$, and $k_{30}$ are the corresponding wave-vectors in vacuum, respectively; $\Delta_2 = 2k_1 - k_2$ is the wave-vector mismatch for the SHG process; and $\Delta_3 = 2k_1 - k_3$ is the wave-vector mismatch for the SFG process. $\chi^{(2)}$ is the natural susceptibility of the waveguide. The spatial modulation of susceptibility is described by the normalized QPM grating function $d(z)$. Here we consider a QPM grating produced by a generalized one-dimensional quasiperiodic optical superlattice as depicted in Fig. 1(b) [17,18,24,25]. The main advantage of such a configuration over the so called "Fibonacci-based structure" is that it can phase match any two arbitrary chosen interactions. The grating function can be expanded in a Fourier series

$$d(z) = \sum_{p,q} d_{pq} e^{iG_{pq}z}, \tag{4}$$

where the summation is over all integers $p$ and $q$ from $-\infty$ to $\infty$. The generalized one-dimensional quasiperiodic QPM grating has two building blocks $A$ and $B$ of length $L_A$ and $L_B$, respectively. Each block contains two inversely polarized ferroelectric domains. Here we assume the length of the positive domains of both blocks $A$ and $B$ are identical, denoted as $S_B$. The front boundary of the $N$th block is given by

$$x_N = L_B \left( N + \alpha + \frac{1}{\rho} \left\lfloor \frac{N}{\sigma} + \beta \right\rfloor \right), \tag{5}$$

where $\alpha, \beta \in (0, 1)$, $\rho$ and $\sigma$ are irrational numbers, and $\lfloor x \rfloor$ represents the largest integer which is smaller than or equal to $x$. The reciprocal vectors are given by

$$G_{pq} = \frac{2\pi}{L_T}\left(p + \frac{q}{\sigma}\right), \tag{6}$$

where $p$ and $q$ are integers. $L_T = L_B[1 + 1/(\rho\sigma)]$ is the average lattice parameter. The Fourier coefficient at $G_{pq}$ is

$$f_{pq} = e^{i\phi_{pq}} \text{sinc} \frac{X_{pq} L_B}{2}, \tag{7}$$

with

$$X_{pq} = \frac{2\pi}{L_T}\left(q - \frac{p}{\rho}\right),$$

$$\phi_{pq} = L_B\left[\alpha G_{pq} - \left(\beta - \frac{1}{2}\right)X_{pq}\right],$$

Let $\alpha = 0, \beta = 1/2$ to simplify the phase item, we obtain the Fourier transform of the generalized quasiperiodic QPM grating as

$$d_{pq} = 2\left(\frac{S_B}{L_T}\right)\exp\left(i\frac{G_{pq}S_B}{2}\right)\text{sinc}\left(\frac{X_{pq}L_B}{2}\right)\text{sinc}\left(\frac{G_{pq}S_B}{2}\right). \tag{8}$$

Figure 2 shows the numerically calculated Fourier spectrum of $d_{pq}$ for $C_b = 0.8$ and four values of $\rho$. As in the following section, the lowest order of the grating wave vector, $d_{01}$ and $d_{10}$, are chosen as the reciprocal vectors quasi-phase-matching for the SHG and SFG processes, respectively.

We introduce the normalized envelopes $A'_1$, $A'_2$, and $A'_3$ according to the relations

$$A_1 = \sqrt{\frac{k_2}{k_1\chi_1\chi_2 x_0^4}}A'_1,\ A_2 = \frac{1}{\chi_1 x_0^2}A'_2,\ A_3 = \sqrt{\frac{k_2}{k_1\chi_1\chi_2 x_0^4}}A'_3, \tag{9}$$

where $\chi_1 = k_{10}^2\chi^{(2)}$, $\chi_2 = k_{20}^2\chi^{(2)}$. By setting $s = x/x_0$ for the propagation coordinate, and $\xi = z/l_d$ for the transverse coordinate, where $x_0$ is the beam width, $l_d = k_1 x_0^2$ is the diffraction length, we obtain a system of coupled equations:

$$i\frac{\partial A'_1}{\partial \xi} - \frac{r}{2}\frac{\partial^2 A'_1}{\partial s^2} + d(\xi)\left(A'_3 A'^*_2 e^{-i\Delta_3\xi} + A'_2 A'^*_1 e^{-i\Delta_2\xi}\right) = 0, \tag{10}$$

$$i\frac{\partial A'_2}{\partial \xi} - \frac{\alpha_2}{2}\frac{\partial^2 A'_2}{\partial s^2} + d(\xi)\left(A'_3 A'^*_1 e^{-i\Delta_3\xi} + \frac{1}{2}A'^2_1 e^{i\Delta_2\xi}\right) = 0, \tag{11}$$

$$i\frac{\partial A'_3}{\partial \xi} - \frac{\alpha_3}{2}\frac{\partial^2 A'_3}{\partial s^2} + d(\xi)\chi A'_2 A'_1 e^{i\Delta_3\xi} = 0, \tag{12}$$

where $r = -1$, $\alpha_2 = -k_1/k_2 \approx -0.5$, $\alpha_3 = -k_1/k_3 \approx -1/3$, and $\chi = 9k_1/k_3 \approx 3$. The wave vector mismatches are transformed to $\Delta_{i\xi} = \Delta_i l_d$, where $i = 2,3$.

Substituting $d(\xi)$ into Eqs. (10), (11), and (12), and making the transformation

$$A'_1 = a_1, A'_2 = a_2 e^{i\beta_2 \xi}, A'_3 = a_3 e^{i\beta_3 \xi}, \tag{13}$$

We obtain the equations

$$i\frac{\partial a_1}{\partial \xi} - \frac{r}{2}\frac{\partial^2 a_1}{\partial s^2} + d_{mn} a_2 a_1^* + d_{gh} a_3 a_2^* + \sum_{m'\neq m \text{ or } n'\neq n} d_{m'n'} a_2 a_1^* e^{i\Delta_{12m'n'}} +$$

$$\sum_{g'\neq g \text{ or } h'\neq h} d_{g'h'} a_3 a_2^* e^{i\Delta_{13g'h'}} = 0$$

$$, \tag{14}$$

$$i\frac{\partial a_2}{\partial \xi} - \beta_2 a_2 - \frac{\alpha_2}{2}\frac{\partial^2 a_2}{\partial s^2} + \frac{1}{2}d_{-m-n} a_1^2 + d_{gh} a_3 a_1^* + \frac{1}{2}\sum_{m'\neq -m \text{ or } n'\neq -n} d_{m'n'} a_1^2 e^{i\Delta_{21m'n'}} +$$

$$\sum_{g'\neq g \text{ or } h'\neq h} d_{g'h'} a_3 a_1^* e^{i\Delta_{13g'h'}} = 0$$

$$, \tag{15}$$

$$i\frac{\partial a_3}{\partial \xi} - \beta_3 a_3 - \frac{\alpha_3}{2}\frac{\partial^2 a_3}{\partial s^2} + \chi d_{-g-h} a_2 a_1 + \chi \sum_{g'\neq -g \text{ or } h'\neq -h} d_{g'h'} a_2 a_1 e^{i\Delta_{31g'h'}} = 0, \tag{16}$$

where $\beta_2 = \Delta_{2\xi} - G_{mn}$ and $\beta_3 = \Delta_{2\xi} - G_{mn} + \Delta_{3\xi} - G_{gh}$ are the effective phase-mismatch parameters. $G_{mn}$ and $G_{gh}$ are the reciprocal vectors chosen in the two nonlinear processes. $\Delta_{12m'n'} = G_{m'n'} - G_{mn}$, $\Delta_{21m'n'} = G_{m'n'} - G_{-m-n}$, $\Delta_{13g'h'} = G_{g'h'} - G_{gh}$ and $\Delta_{31g'h'} = G_{g'h'} - G_{-g-h}$ are parameters representing the normalized phase-mismatches for grating wave vectors of different orders.

We consider here a typical QPM grating for which the domain length is much smaller than the diffraction length. Following the method introduced by Clausen et al [12], we expand the field $a_1$, $a_2$, and $a_3$ in Fourier series

$$a_1 = u_{0,0} + \sum_{t_1 \neq 0 \text{ or } q_1 \neq 0} u_{t_1,q_1} e^{iM_1\xi},$$

$$a_2 = v_{0,0} + \sum_{t_2 \neq 0 \text{ or } q_2 \neq 0} v_{t_2,q_2} e^{iM_2\xi},$$

$$a_3 = w_0 + \sum_{q_3 \neq 0} w_{q_3} e^{iM_3\xi}, \tag{17}$$

where $u_{t_1,q_1}(s,\xi)$, $v_{t_2,q_2}(s,\xi)$ and $w_{q_3}(s,\xi)$ are assumed to vary slowly compared with

$\exp(iG_{pq}\xi)$, and $M_{1\xi} = t_1 G_{mn} + q_1 G_{gh}$, $M_{2\xi} = t_2 G_{mn} + q_2 G_{gh}$, $M_{3\xi} = q_3 G_{gh}$. Following the reasoning of Ref. [12], we assume that the higher harmonics are smaller, compared to the averages $u_{0,0}$, $v_{0,0}$, and $w_0$. Taking into account only the lowest order terms in the equations for the harmonics, we then derive the following relations:

$$u_{t_1,q_1} = \frac{\frac{d_{(t_1+1)m+q_1g,}v_{0,0}u_{0,0}^* + d_{t_1m+(q_1+1)g,}w_0 v_{0,0}^*}{(t_1+1)n+q_1h} + \frac{d_{t_1m+(q_1+1)g,}w_0 v_{0,0}^*}{t_1n+(q_1+1)h}}{M_{1\xi}},$$

$$v_{t_2,q_2} = \frac{\frac{\frac{1}{2}d_{(t_2-1)m+q_2g,}u_{0,0}^2 + d_{t_2m+(q_2+1)g,}w_0 u_{0,0}^*}{(t_2-1)n+q_2h} + \frac{d_{t_2m+(q_2+1)g,}w_0 u_{0,0}^*}{t_2n+(q_2+1)h}}{M_{2\xi}+\beta_2},$$

$$w_{q_3} = \frac{\chi d_{(q_3-1)g,(q_3-1)h} u_{0,0} v_{0,0}}{M_{3\xi}+\beta_3}, \tag{18}$$

Inserting the harmonics (18) into the corresponding equations for $u_{0,0}$, $v_{0,0}$, and $w_0$, and taking only lowest order perturbations into account, we arrive at the average equations

$$i\frac{\partial u_{0,0}}{\partial \xi} - \frac{r}{2}\frac{\partial^2 u_{0,0}}{\partial s^2} + d_{mn}v_{0,0}u_{0,0}^* + d_{gh}w_0 v_{0,0}^* + \frac{1}{2}\gamma_{11}|u_{0,0}|^2 u_{0,0} + \gamma_{12}|v_{0,0}|^2 u_{0,0} + \gamma_{13}|w_0|^2 u_{0,0} +$$

$$\frac{3}{2}\rho_{13}w_0(u_{0,0}^*)^2 + \rho_{123}v_{0,0}^2 w_0^* = 0$$

, (19)

$$i\frac{\partial v_{0,0}}{\partial \xi} - \beta_2 v_{0,0} - \frac{\alpha_2}{2}\frac{\partial^2 v_{0,0}}{\partial s^2} + \frac{1}{2}d_{-m-n}u_{0,0}^2 + d_{gh}w_0 u_{0,0}^* + \gamma_{21}|u_{0,0}|^2 v_{0,0} + \gamma_{23}|w_0|^2 v_{0,0} +$$

$$2\rho_{231}u_{0,0}w_0 v_{0,0}^* = 0$$

, (20)

$$i\frac{\partial w_0}{\partial \xi} - \beta_3 w_0 - \frac{\alpha_3}{2}\frac{\partial^2 w_0}{\partial s^2} + \chi d_{-g-h}u_{0,0}v_{0,0} + \chi\gamma_{31}|u_{0,0}|^2 w_0 + \chi\gamma_{32}|v_{0,0}|^2 w_0 + \frac{\chi}{2}\rho_{31}u_{0,0}^3 +$$

$$\chi\rho_{312}v_{0,0}^2 u_{0,0}^* = 0$$

, (21)

where

$$\gamma_{11} = \sum_{t_2 \neq 0 \text{ or } q_2 \neq 0} \frac{d_{-(t_2-1)m-q_2g,}d_{(t_2-1)m+q_2g,}}{-(t_2-1)n-q_2h \quad (t_2-1)n+q_2h}}{M_2},$$

$$\gamma_{12} = \sum_{t_1 \neq 0 \text{ or } q_1 \neq 0} \frac{d_{(t_1+1)m+q_1g,}d_{(t_1+1)m+q_1g,}^*}{(t_1+1)n+q_1h \quad (t_1+1)n+q_1h}}{M_1} + \chi \sum_{q_3 \neq 0} \frac{d_{-(q_3-1)g,}d_{(q_3-1)g,}}{-(q_3-1)h \quad (q_3-1)h}}{M_3},$$

$$\gamma_{21} = \sum_{t_1 \neq 0 \text{ or } q_1 \neq 0} \frac{d_{(t_1+1)m+q_1 g, (t_1+1)n+q_1 h} d^*_{-(t_1+1)m-q_1 g, -(t_1+1)n-q_1 h}}{M_1} + \chi \sum_{q_3 \neq 0} \frac{d_{-(q_3-1)g, -(q_3-1)h} d_{(q_3-1)g, (q_3-1)h}}{M_3},$$

$$\gamma_{13} = \sum_{t_2 \neq 0 \text{ or } q_2 \neq 0} \frac{d_{t_2 m+(q_2+1)g, t_2 n+(q_2+1)h} d^*_{t_2 m+(q_2+1)g, t_2 n+(q_2+1)h}}{M_2},$$

$$\gamma_{31} = \sum_{t_2 \neq 0 \text{ or } q_2 \neq 0} \frac{d_{t_2 m+(q_2+1)g, t_2 n+(q_2+1)h} d_{-t_2 m-(q_2+1)g, -t_2 n-(q_2+1)h}}{M_2},$$

$$\gamma_{23} = \sum_{t_1 \neq 0 \text{ or } q_1 \neq 0} \frac{d_{t_1 m+(q_1+1)g, t_1 n+(q_1+1)h} d^*_{t_1 m+(q_1+1)g, t_1 n+(q_1+1)h}}{M_1},$$

$$\gamma_{32} = \sum_{t_1 \neq 0 \text{ or } q_1 \neq 0} \frac{d_{t_1 m+(q_1+1)g, t_1 n+(q_1+1)h} d_{-t_1 m-(q_1+1)g, -t_1 n-(q_1+1)h}}{M_1},$$

$$\rho_{13} = \frac{2}{3} \left( \frac{1}{2} \sum_{t_2 \neq 0 \text{ or } q_2 \neq 0} \frac{d_{t_2 m+(q_2+1)g, t_2 n+(q_2+1)h} d^*_{(t_2-1)m+q_2 g, (t_2-1)n+q_2 h}}{M_2} + \sum_{t_2 \neq 0 \text{ or } q_2 \neq 0} \frac{d_{t_2 m+(q_2+1)g, t_2 n+(q_2+1)h} d_{-(t_2-1)m-q_2 g, -(t_2-1)n-q_2 h}}{M_2} \right),$$

$$\rho_{31} = \sum_{t_2 \neq 0 \text{ or } q_2 \neq 0} \frac{d_{-t_2 m-(q_2+1)g, -t_2 n-(q_2+1)h} d_{(t_2-1)m+q_2 g, (t_2-1)n+q_2 h}}{M_2},$$

$$\rho_{123} = \sum_{t_1 \neq 0 \text{ or } q_1 \neq 0} \frac{d_{(t_1+1)m+q_1 g, (t_1+1)n+q_1 h} d^*_{t_1 m+(q_1+1)g, t_1 n+(q_1+1)h}}{M_1},$$

$$\rho_{231} = \frac{1}{2} \left( \sum_{t_1 \neq 0 \text{ or } q_1 \neq 0} \frac{d_{t_1 m+(q_1+1)g, t_1 n+(q_1+1)h} d^*_{(t_1+1)m+q_1 g, (t_1+1)n+q_1 h}}{M_1} + \sum_{t_1 \neq 0 \text{ or } q_1 \neq 0} \frac{d_{t_1 m+(q_1+1)g, t_1 n+(q_1+1)h} d_{-(t_1+1)m-q_1 g, -(t_1+1)n-q_1 h}}{M_1} \right),$$

$$\rho_{312} = \sum_{t_1 \neq 0 \text{ or } q_1 \neq 0} \frac{d_{(t_1+1)m+q_1g,\,-t_1m-(q_1+1)g,}^{\phantom{(t_1+1)m+q_1g,}}\, d_{-t_1m-(q_1+1)g,}^{\phantom{(t_1+1)m+q_1g,}}}{M_1},$$

By using $d_{-p-q} = d_{pq}^*$ obtained from expression (8), we arrive at the results

$$\gamma_{12} = \gamma_{21},\ \gamma_{13} = \gamma_{31},\ \gamma_{23} = \gamma_{32},\ \rho_{13} = \rho_{31}^*,\ \rho_{123} = \rho_{231}^* = \rho_{312}.$$

Similar to Clausen's results, the QPM grating introduces the effective cubic nonlinearity in the form of self- and cross-phase modulation terms as a result of the SHG process. However, there is an important difference: due the SFG process, a third harmonic generation and four-wave mixing terms appear for the formation of multicolor solitons.

To obtain a great insight into the characteristic of the quasiperiodic multicolor solitons, we take a consideration of the integrals of the wave evolution. Here we shall make use of three known integrals which can be readily obtained from Noether's theorem, or directly from the governing equations, namely, the total beam power or energy flow given by the Manley-Rowe relation

$$I = \int \left( |u_{0,0}|^2 + 2|v_{0,0}|^2 + \frac{3}{\chi}|w_0|^2 \right) ds, \tag{22}$$

the Hamiltonian or field energy

$$H = -\frac{1}{2}\int \left( r\left|\frac{\partial u_{0,0}}{\partial s}\right|^2 + \alpha_2\left|\frac{\partial v_{0,0}}{\partial s}\right|^2 + \frac{\alpha_3}{\chi}\left|\frac{\partial w_0}{\partial s}\right|^2 - 2\beta_2|v_{0,0}|^2 - \frac{2\beta_3}{\chi}|w_0|^2 + \left(d_{mn}(u_{0,0}^*)^2 v_{0,0} + d_{-m-n}v_{0,0}^* u_{0,0}^2\right) + 2\left(d_{gh}u_{0,0}^* v_{0,0}^* w_0 + d_{-g-h}u_{0,0}v_{0,0}w_0^*\right) + \frac{\gamma_{11}}{2}|u_{0,0}|^4 + 2\gamma_{12}|u_{0,0}|^2|v_{0,0}|^2 + 2\gamma_{13}|u_{0,0}|^2|w_0|^2 + 2\gamma_{23}|v_{0,0}|^2|w_0|^2 + \left(\rho_{13}(u_{0,0}^*)^3 w_0 + \rho_{31}u_{0,0}^3 w_0^*\right) + 2\left(\rho_{123}u_{0,0}^* v_{0,0}^2 w_0^* + \rho_{231}u_{0,0}(v_{0,0}^*)^2 w_0\right) \right) ds$$

, (23)

and the total transverse beam momentum

$$J = \frac{i}{2}\int \left( \left(u_{0,0}^*\frac{\partial u_{0,0}}{\partial s} - u_{0,0}\frac{\partial u_{0,0}^*}{\partial s}\right) + \left(v_{0,0}^*\frac{\partial v_{0,0}}{\partial s} - v_{0,0}\frac{\partial v_{0,0}^*}{\partial s}\right) + \frac{1}{\chi}\left(w_0^*\frac{\partial w_0}{\partial s} - w_0\frac{\partial w_0^*}{\partial s}\right) \right) ds, \tag{24}$$

The number of integrals of motion of the evolution system is closely related to the number of internal soliton parameters of the corresponding soliton families which, in turn, are related to the existing symmetries.

## III. MULTICOLOR SOLITONS

We look for stationary solutions to Eqs. (19), (20), and (21) in the form

$$u_{0,0}(s,\xi) = u(s)e^{i\kappa_1\xi},\ v_{0,0}(s,\xi) = v(s)e^{i2\kappa_1\xi},\ w_0(s,\xi) = w(s)e^{i3\kappa_1\xi}, \quad (25)$$

Substitution of (25) into Eqs. (19), (20), and (21) yields the set of nonlinear ordinary differential equations

$$-\frac{r}{2}\frac{\partial^2 u}{\partial s^2} - \kappa_1 u + d_{mn}u^*v + d_{gh}v^*w + \frac{1}{2}\gamma_{11}|u|^2 u + \gamma_{12}|v|^2 u + \gamma_{13}|w|^2 u + \frac{3}{2}\rho_{13}(u^*)^2 w + \rho_{123}v^2 w^* = 0$$

, (26)

$$-\frac{\alpha_2}{2}\frac{\partial^2 v}{\partial s^2} - (2\kappa_1 + \beta_2)v + \frac{1}{2}d_{-m-n}u^2 + d_{gh}u^*w + \gamma_{21}|u|^2 v + \gamma_{23}|w|^2 v + 2\rho_{231}uv^*w = 0, \quad (27)$$

$$-\frac{\alpha_3}{2}\frac{\partial^2 w}{\partial s^2} - (3\kappa_1 + \beta_3)w + \chi d_{-g-h}uv + \chi\gamma_{31}|u|^2 w + \chi\gamma_{32}|v|^2 w + \frac{\chi}{2}\rho_{31}u^3 + \chi\rho_{312}u^*v^2 = 0,$$

(28)

In order to simplify the Eqs. (26), (27), and (28) into a real form, we introduce

$$u = \bar{u}e^{i\theta_u},\ v = \bar{v}e^{i\theta_v},\ w = \bar{w}e^{i\theta_w},$$

$$d_{mn} = \bar{d}_{mn}e^{i\theta_{mn}},\ d_{gh} = \bar{d}_{gh}e^{i\theta_{gh}},\ \rho_{13} = \bar{\rho}_{13}e^{i\theta_{13}},\ \rho_{123} = \bar{\rho}_{123}e^{i\theta_{123}}, \quad (29)$$

where $\theta_u$ and $\theta_{mn}$ are phase angles of $u$ and $d_{mn}$, respectively, and $\theta_v = 2\theta_u - \theta_{mn}$, $\theta_w = \theta_u + \theta_v - \theta_{mn}$, $\theta_{13} = \theta_{mn} + \theta_{gh}$, and $\theta_{123} = \theta_{mn} - \theta_{gh}$. The real and localized profiles $\bar{u}(s)$, $\bar{v}(s)$, and $\bar{w}(s)$ are determined by the set of ordinary differential equations

$$-\frac{r}{2}\frac{\partial^2 \bar{u}}{\partial s^2} - \kappa_1\bar{u} + \bar{d}_{mn}\bar{u}\bar{v} + \bar{d}_{gh}\bar{v}\bar{w} + \frac{1}{2}\gamma_{11}\bar{u}^3 + \gamma_{12}\bar{v}^2\bar{u} + \gamma_{13}\bar{w}^2\bar{u} + \frac{3}{2}\bar{\rho}_{13}\bar{u}^2\bar{w} + \bar{\rho}_{123}\bar{v}^2\bar{w} = 0,$$

(30)

$$-\frac{\alpha_2}{2}\frac{\partial^2 \bar{v}}{\partial s^2} - (2\kappa_1 + \beta_2)\bar{v} + \frac{1}{2}d_{-m-n}\bar{u}^2 + \bar{d}_{gh}\bar{u}\bar{w} + \gamma_{21}\bar{u}^2\bar{v} + \gamma_{23}\bar{w}^2\bar{v} + 2\bar{\rho}_{231}\bar{u}\bar{v}\bar{w} = 0, \quad (31)$$

$$-\frac{\alpha_3}{2}\frac{\partial^2 \bar{w}}{\partial s^2} - (3\kappa_1 + \beta_3)\bar{w} + \chi d_{-g-h}\bar{u}\bar{v} + \chi\gamma_{31}\bar{u}^2\bar{w} + \chi\gamma_{32}\bar{v}^2\bar{w} + \frac{\chi}{2}\rho_{31}\bar{u}^3 + \chi\bar{\rho}_{312}\bar{u}\bar{v}^2 = 0, \quad (32)$$

Analysis shows that localized solutions (29) exist only for positive values of the wave number $\kappa_1$, satisfying $\kappa_1 > \max(0, -\beta_2/2, -\beta_3/2)$.

We solve Eqs. (30), (31), and (32) numerically using a band-matrix method to deal with the two-point boundary value problem for the unknown functions $\bar{u}(s), \bar{v}(s)$, and $\bar{w}(s)$. Figure 3 shows the characteristic profiles of the solutions for $C_b = 0.8$ and four values of $\rho$. The peak amplitude of the TH varies drastically while that of the SH remains nearly constant for different values of $\rho$. In the calculation, the Fourier transforms of the lowest order of the grating wave vectors, $d_{01}$ and $d_{10}$, are chosen as the reciprocal vectors quasi-phase-matching for the SHG and SFG processes, respectively. As shown in Fig. 2, for the TH, which is determined by the SFG process, its peak amplitude is mainly influenced by the value of $d_{10}$; while for the SH, both the SHG and SFG processes play an important role, making both $d_{01}$ and $d_{10}$ equally important.

In order to test our asymptotic results, we use a quasiperiodic QPM soliton as the initial condition in Eqs. (1), (2), and (3), which we solve numerically for the quasiperiodic grating shown in Fig. 1. The results for $\rho = 1.5$ and $C_b = 0.9$ are plotted in Fig. 4 and show clearly that the generated three component soliton propagates quasiperiodically along z. As a matter of fact, after the initial transient, its amplitude oscillates in phase with the quasiperiodic QPM modulation $d(z)$. This is illustrated in Fig. 4(d), where we show in more detail the peak intensities in the asymptotic regime of the evolution.

Since in real experiments one cannot launch into a QPM grating beams with spatial shapes that rigorously match those of specific multicolor solitons, it is very important to study whether the quasiperiodic multicolor solitons described here can be excited from the beams that are more accessible experimentally, that is, the Gaussian beams. We start by launching the FF and its SH in the form of Gaussian beams, while setting the TH equal to 0

$$u(s,0) = A_u e^{-s^2/w_u^2}, \quad v(s,0) = A_v e^{-s^2/w_v^2}, \quad w(s,0) = 0. \tag{33}$$

First, we study the small amplitude limit when both a weak FF and SH are injected with amplitude

$A_u = 5, A_v = 5$. We obtain that the TH is excited, but fluctuations and diffraction prevail so that all the FF, SH and TH eventually spread out. Then we investigate the evolution of a strong input FF and SH beams with $A_u = 20, A_v = 20$, and its corresponding TH. Figure 5 shows the results corresponding excitation of multicolor solitons with the seeded FF and SH. which leads to self-focusing and mutual self-trapping of the three fields. Due to the strong self-trapping effect of the nonlinearity, the soliton remains nicely localized for more than 20 diffraction lengths. In typical experiments a stable spatial soliton is said to be observed if it remains localized for about 5–10 diffraction lengths. Furthermore, similar as shown in Fig. 4(d), the peak intensity of these three components oscillate in phase with the quasiperiodic QPM modulation $d(z)$. Thus, for all practical purposes, the soliton generated here is dynamically stable.

**IV. STABILITY AND WAVE COLLAPES OF THE MULTICOLOR SOLITONS**

**A. The stability of the multicolor solitons**

We now prove the stability of the coupled-soliton solution. By substituting expressions (25) into Eqs. (22), (23), and (24), we rewrite these equations in new forms

$$I = I_1 + I_2 + I_3 = \int \left(|u|^2 + 2|v|^2 + \frac{3}{\chi}|w|^2\right) ds, \tag{34}$$

$$H = \int \left(\beta_2 |v|^2 + \frac{\beta_3}{\chi}|w|^2 - \frac{r}{2}\left|\frac{\partial u}{\partial s}\right|^2 - \frac{\alpha_2}{2}\left|\frac{\partial v}{\partial s}\right|^2 - \frac{\alpha_3}{2\chi}\left|\frac{\partial w}{\partial s}\right|^2 - \text{Re}(d_{mn}(u^*)^2 v) - 2\text{Re}(d_{gh}u^*v^*w) - \frac{\gamma_{11}}{4}|u|^4 - \gamma_{12}|u|^2|v|^2 - \gamma_{13}|u|^2|w|^2 - \gamma_{23}|v|^2|w|^2 - \text{Re}(\rho_{13}(u^*)^3 w) - 2\text{Re}(\rho_{123}u^*v^2w^*)\right) ds,$$

$$\tag{35}$$

$$J = \frac{i}{2}\int \left(\left(u^*\frac{\partial u}{\partial s} - u\frac{\partial u^*}{\partial s}\right) + \left(v^*\frac{\partial v}{\partial s} - v\frac{\partial v^*}{\partial s}\right) + \frac{1}{\chi}\left(w^*\frac{\partial w}{\partial s} - w\frac{\partial w^*}{\partial s}\right)\right) ds. \tag{36}$$

We start by employing the Lyapunov-type analysis based on proving the multicolor soliton realize a minimum of Hamiltonian (35) for fixed values of invariants (34) and (36). We multiply Eq. (26) by $(u^*)$, on the one hand, and by $(s\,\partial u^*/\partial s)$, on the other hand, then integrate in space both of the resulting equations. We next repeat the previous operations in Eqs. (27) and (28) by formally replacing $u \to v, w$.

By introducing

$$L_1 = \frac{1}{4}\int \left(r\left|\frac{\partial u}{\partial s}\right|^2 + \alpha_2\left|\frac{\partial v}{\partial s}\right|^2 + \frac{\alpha_3}{\chi}\left|\frac{\partial w}{\partial s}\right|^2\right)ds,$$

$$L_2 = \int \left(\frac{\gamma_{11}}{8}|u|^4 + \frac{\gamma_{12}}{2}|u|^2|v|^2 + \frac{\gamma_{13}}{2}|u|^2|w|^2 + \frac{\gamma_{23}}{2}|v|^2|w|^2 + \frac{1}{2}\text{Re}(\rho_{13}(u^*)^3 w) + \text{Re}(\rho_{123}u^*v^2w^*)\right)ds$$

,

$$L_3 = \text{Re}\int (d_{mn}(u^*)^2 v + d_{gh}u^*v^*w)ds, \tag{37}$$

a simple combination of the space-integrated results yields

$$2L_1 - \kappa_1 I - \frac{\beta_2}{2}I_2 - \frac{\beta_3}{3}I_3 + 4L_2 + 3L_3 = 0, \tag{38}$$

$$L_1 + \frac{\kappa_1}{2}I + \frac{\beta_2}{4}I_2 + \frac{\beta_3}{6}I_3 - L_2 - L_3 = 0, \tag{39}$$

where $I_2 \equiv 2\int |v|^2 ds$ and $I_3 \equiv \frac{3}{\chi}\int |w|^2 ds$ are the second and third harmonic components of the total beam power, respectively. By a simple combination of Eqs. (38) and (39), we obtain

$$4L_1 + 2L_2 + L_3 = 0. \tag{40}$$

The Hamiltonian (35) can be rewritten as

$$H = \frac{\beta_2}{2}I_2 + \frac{\beta_3}{2}I_3 - 2L_1 - 2L_2 - 2L_3, \tag{41}$$

Substituting expression (29) into the Hamiltonian (35), we now prove the stability of the coupled-soliton solution by arguing that the functional

$$S = H - \frac{\beta_2}{2}I_2 - \frac{\beta_3}{3}I_3 = -\int \left(\frac{r}{2}\left(\frac{\partial \bar{u}}{\partial s}\right)^2 + \frac{\alpha_2}{2}\left(\frac{\partial \bar{v}}{\partial s}\right)^2 + \frac{\alpha_3}{2\chi}\left(\frac{\partial \bar{w}}{\partial s}\right)^2 + \bar{d}_{mn}\bar{u}^2\bar{v} + 2\bar{d}_{gh}\bar{u}\bar{v}\bar{w} + \frac{\gamma_{11}}{4}\bar{u}^4 + \gamma_{12}\bar{u}^2\bar{v}^2 + \gamma_{13}\bar{u}^2\bar{w}^2 + \gamma_{23}\bar{v}^2\bar{w}^2 + \bar{\rho}_{13}\bar{u}^3\bar{w} + 2\bar{\rho}_{123}\bar{u}\bar{v}^2\bar{w}\right)ds$$

, (42)

can be viewed as a Lyapunov function that remains bounded from below, and whose bound admits for a couple of fixed invariants $I_2$, $I_3$ a global minimum reached on the coupled-soliton family. We make use of the Schwarz and Sobolev inequalities,

$$\int \bar{u}^2 \bar{v} ds \leq C_{uv} (\int \bar{u}^2 ds)^{\frac{3}{4}} \left[\int \left(\frac{\partial \bar{u}}{\partial s}\right)^2 ds\right]^{\frac{1}{4}} (\int \bar{v}^2 ds)^{\frac{1}{2}}, \quad (43)$$

where $C_{uv}$ denotes a positive constant. By means of this inequality, $H$ is bounded from below as follows:

$$H - \frac{\beta_2}{2} I_2 - \frac{\beta_3}{3} I_3 \geq \bar{U}_s \left[\left(-\frac{r}{2}\right) - (\bar{d}_{mn} + \bar{d}_{gh}) C_{uv} I_1^{\frac{3}{4}} \left(\frac{I_2}{2}\right)^{\frac{1}{2}} \bar{U}_s^{-\frac{3}{4}} - N_1 C_u^2 I_1^{\frac{3}{2}} \bar{U}_s^{-\frac{1}{2}}\right] + \bar{V}_s \left[\left(-\frac{\alpha_2}{2}\right) - N_2 C_v^2 \left(\frac{I_2}{2}\right)^{\frac{3}{2}} \bar{V}_s^{-\frac{1}{2}}\right] + \bar{W}_s \left[\left(-\frac{\alpha_3}{2\chi}\right) - \bar{d}_{gh} C_{vw} \left(\frac{\chi I_3}{3}\right)^{\frac{3}{4}} \left(\frac{I_2}{2}\right)^{\frac{1}{2}} \bar{W}_s^{-\frac{3}{4}} - N_3 C_w^2 \left(\frac{\chi I_3}{3}\right)^{\frac{3}{2}} \bar{W}_s^{-\frac{1}{2}}\right]$$

, (44)

where $\bar{U}_s = \int (\partial \bar{u}/\partial s)^2 ds$ , $\bar{V}_s = \int (\partial \bar{v}/\partial s)^2 ds$ , and $\bar{W}_s = \int (\partial \bar{w}/\partial s)^2 ds$ .

$N_1 = \gamma_{11}/4 + \gamma_{12}/2 + \gamma_{13}/2 + 3\bar{\rho}_{13}/4 + \bar{\rho}_{123}/2$ , $N_2 = \gamma_{12}/2 + \gamma_{23}/2 - \bar{\rho}_{123}$ ,

$N_3 = \gamma_{13}/2 + \gamma_{23}/2 + \bar{\rho}_{13}/4 + \bar{\rho}_{123}/2$ are positive constants. The integral (42) is indeed bounded from below by the functionals of $\bar{U}_s$, $\bar{V}_s$, and $\bar{W}_s$ reaching a global minimum, as seen from the right-hand side of the inequality (44). The latter estimate indicates that Eqs. (19), (20) and (21) admit some fixed-point (stationary) solutions that are stable. In fact, the important integral, from which the stability of the stationary solutions follows, is the Hamiltonian $H$ that exhibits a strict minimum. To examine which kind of fixed-point solutions can realize this minimum of $H$, we use the property according to which functional $S$ admits a single minimum. Following the standard procedure reviewed in [26] and [27], the minimum of $S$ is identified by using the scale transformations

$$u(s,\xi) \to a^{-\frac{1}{2}} u(s/a,\xi), v(s,\xi) \to a^{-\frac{1}{2}} v(s/a,\xi), w(s,\xi) \to a^{-\frac{1}{2}} w(s/a,\xi) \quad (45)$$

that preserve the $L^2$ norms attached to each wave $u$, $v$ and $w$ in expression (23). Here, $a$ denotes a constant parameter playing the role of a Lagrange multiplier that only affects the energy integral when one inserts (45) into $S$, leading to

$$S_a = -a^{-2}L_1 - 2a^{-\frac{1}{2}}L_3 - 2a^{-1}L_2. \tag{46}$$

Differentiating $S$, with respect to the parameter $a$ constrained on the value $a = 1$ then yields the minimum of $S$. By doing so, one deduces that the latter functional admits a single minimum reached when the solutions satisfy the relation $4L_1 + 2L_2 + L_3 = 0$, which is nothing else but the relation (40) realized by the ground-state solutions. Hence, as inferred from the previous variational problem $\delta S = 0$, $S$ contains a stable fixed point, on which its minimum is reached and which corresponds to the coupled-soliton solutions. This minimum also corresponds to the minimum of $H$, so that we now obtain the inequality

$$H \geq H_{GS}, \tag{47}$$

**B. Absence of wave collapse**

In order to investigate the various dynamical aspects of the coupled waves $(u, v, w)$, we construct a so-called "virial" identity [28,29], in analogy with the standard result of the NSE, consisting in the double derivative with respect to the longitudinal distance $\xi$ of the mean square radius

$$I_v(\xi) = \int s^2 \left( |u|^2 + 2|v|^2 + \frac{3}{\chi}|w|^2 \right) ds, \tag{48}$$

We first multiply Eq. (26) by $(s^2 u^*)$, Eq. (27) by $(s^2 v^*)$ and Eq. (28) by $(s^2 w^*)$ and sum up the imaginary part of the space integrated results to get

$$\frac{\partial I_v}{\partial \xi} = -2\mathrm{Im}\int s\left( r u^* \frac{\partial u}{\partial s} + 2\alpha_2 v^* \frac{\partial v}{\partial s} + \frac{3}{\chi}\alpha_3 w^* \frac{\partial w}{\partial s} \right), \tag{49}$$

Then we multiply Eq. (26) by $(s\, \partial u^*/\partial s)$ and integrate the real part of the result to obtain after a few integrations by parts

$$\frac{\partial}{\partial \xi} \mathrm{Im} \int s u^* \frac{\partial u}{\partial s} ds = -r \int \left|\frac{\partial u}{\partial s}\right|^2 ds + \mathrm{Re}\left[d_{mn} \int s(u^*)^2 \frac{\partial v}{\partial s} ds\right] + \mathrm{Re}\left[d_{gh} \int \left( u^* v^* w + \right.\right.$$
$$\left.\left. 2su^* w \frac{\partial v^*}{\partial s} + 2su^* v^* \frac{\partial w}{\partial s} \right)ds \right] - \frac{\gamma_{11}}{4} \int |u|^4 ds + \gamma_{12} \int s|u|^2 \frac{\partial |v|^2}{\partial s} ds + \gamma_{13} \int s|u|^2 \frac{\partial |w|^2}{\partial s} ds +$$
$$\mathrm{Re}\left\{ \rho_{13} \int \left[ s(u^*)^3 \frac{\partial w}{\partial s} - \frac{1}{2}(u^*)^3 w \right] ds \right\} + \mathrm{Re}\left[ \rho_{123} \int \left( u^* v^2 w^* + 2su^* w \frac{\partial v^2}{\partial s} + 2su^* v^2 \frac{\partial w^*}{\partial s} \right) ds \right]$$

,

(50)

Next, repeating the same procedure on Eqs. (27) and (28), we multiply the former by $(s\, \partial v^*/\partial s)$ and the latter by $(s\, \partial w^*/\partial s)$ to find

$$\frac{\partial}{\partial \xi} \mathrm{Im} \int s v^* \frac{\partial v}{\partial s} ds =$$

$$-\alpha_2 \int \left|\frac{\partial v}{\partial s}\right|^2 ds - \mathrm{Re}\left[d_{-m-n} \int \left(s u^2 \frac{\partial v^*}{\partial s} + \frac{1}{2} u^2 v^*\right) ds\right] - \mathrm{Re}\left[d_{gh} \int \left(u^* v^* w + 2 s u^* w \frac{\partial v^*}{\partial s}\right) ds\right] -$$

$$\gamma_{21} \int \left(|u|^2 |v|^2 + s|u|^2 \frac{\partial |v|^2}{\partial s}\right) ds + \gamma_{23} \int s|v|^2 \frac{\partial |w|^2}{\partial s} ds -$$

$$2\mathrm{Re}\left\{\rho_{231} \int \left[u(v^*)^2 w + s u w \frac{\partial (v^*)^2}{\partial s}\right] ds\right\},$$

(51)

$$\frac{1}{\chi} \frac{\partial}{\partial \xi} \mathrm{Im} \int s w^* \frac{\partial w}{\partial s} ds =$$

$$-\frac{\alpha_3}{\chi} \int \left|\frac{\partial w}{\partial s}\right|^2 ds - \mathrm{Re}\left[d_{-g-h} \int \left(2 s u v \frac{\partial w^*}{\partial s} + u v w^*\right) ds\right] - \gamma_{31} \int \left(|u|^2 |w|^2 + s|u|^2 \frac{\partial |w|^2}{\partial s}\right) ds -$$

$$\gamma_{32} \int \left(s|v|^2 \frac{\partial |w|^2}{\partial s} + |v|^2 |w|^2\right) ds - \mathrm{Re}\left\{\rho_{312} \int \left[u^* v^2 w^* + 2 s u^* v^2 \frac{\partial w}{\partial s}\right] ds\right\} -$$

$$\mathrm{Re}\left[\rho_{31} \int \left[\frac{1}{2} u^3 w^* + s u^3 \frac{\partial w^*}{\partial s}\right] ds\right]$$

.
(52)

Combining Eqs. (50), (51), and (52) into the $\xi$ derivative of Eq. (49) finally yields the virial identity

$$\frac{\partial^2 I_v}{\partial \xi^2} =$$

$$2\int \left|\frac{\partial u}{\partial s}\right|^2 ds + \int \left|\frac{\partial v}{\partial s}\right|^2 ds + \frac{2}{9}\int \left|\frac{\partial w}{\partial s}\right|^2 ds - \mathrm{Re}(d_{mn} \int (u^*)^2 v ds) - 2\mathrm{Re}(d_{gh} \int u^* v^* w ds) -$$

$$\frac{\gamma_{11}}{2} \int |u|^4 ds - 2\gamma_{12} \int |u|^2 |v|^2 ds - 2\gamma_{13} \int |u|^2 |w|^2 ds - 2\gamma_{23} \int |v|^2 |w|^2 ds -$$

$$2\mathrm{Re}(\rho_{13} \int (u^*)^3 w ds) - 4\mathrm{Re}(\rho_{123} \int u^* v^2 w^* ds),$$

(53)

where we set $r = -1$, $\alpha_2 = -1/2$, $\alpha_3 = -1/3$, and $\chi = 3$. Eq. (53) can also be written, using the definition (23), in the alternative forms

$$\frac{\partial^2 I_v}{\partial \xi^2} = 2\left(H - \beta_2 \int |\frac{\partial v}{\partial s}|^2 ds - \frac{\beta_3}{\chi}\int |\frac{\partial w}{\partial s}|^2 ds\right) + \text{Re}[d_{mn}\int(u^*)^2 v ds] + 2\text{Re}(d_{gh}\int u^* v^* w ds) +$$
$$\frac{1}{2}\int |\frac{\partial u}{\partial s}|^2 ds + \frac{1}{4}\int |\frac{\partial v}{\partial s}|^2 ds + \frac{1}{19}\int |\frac{\partial w}{\partial s}|^2 ds$$

, (54)

From expression (54), we recognize the typical functional $(H - \beta_2 \int |\partial v/\partial s|^2 ds - \beta_3/\chi \int |\partial w/\partial s|^2 ds)$. Let us suppose that a collapse of solutions $u(s,\xi)$, $v(s,\xi)$ and $w(s,\xi)$ may occur in the sense that $I_v(\xi) \to 0$ at a finite $\xi = \xi_c$ with, e.g., $\partial^2 I_v/\partial \xi^2 < 0$. Then necessarily both the integrals $\int s^2|u|^2 ds$, $\int s^2|v|^2 ds$ and $\int s^2|w|^2 ds$ have to vanish separately and simultaneously as $\xi \to \xi_c$. Employing the Schwarz inequality after a simple integration by parts, we obtain

$$(\int |g|^2 ds)^2 \le 4\int s^2 |g|^2 ds \int |\frac{\partial g}{\partial s}|^2 ds, \qquad (55)$$

applied to any $L^2$-integrable function $g$, so that the previous assumption should imply that both the gradient norms $\int |\partial u/\partial s|^2 ds$, $\int |\partial v/\partial s|^2 ds$ and $\int |\partial w/\partial s|^2 ds$ diverge as $\xi \to \xi_c$. By virtue of the constancy of $H$ and since the finiteness of $I$ ensures that the two masses $I_2 \equiv 2\int |v_{0,0}|^2 ds$ and $I_3 \equiv \frac{3}{\chi}\int |w_0|^2 ds$ remain bounded, the quantity $\partial^2 I_v/\partial \xi^2$ should thus diverge in the vicinity of the collapse focus $\xi_c$, hence predicting the spreading of both wave forms, which contradicts the starting hypothesis. Therefore, $\xi$-dependent solutions $u$, $v$ and $w$ can never collapse at any finite distance $\xi$, and will be expected to exist globally for every $\xi$ by keeping a bounded gradient norm.

**C. The stability criterion of the solitons**

Next we derive the condition of marginal stability of the family of solutions by applying a multiscale asymptotic method [30-33]. Let $X_0 = (u_{ore}, u_{oim}, v_{ore}, v_{oim}, w_{ore}, w_{oim})^T$ be the column vector formed with the stationary quasiperiodic solitons

$u_0(s,\xi) = u_{ore}(s) + iu_{oim}(s)$, $v_0(s,\xi) = v_{ore}(s) + iv_{oim}(s)$, $w_0(s,\xi) = w_{ore}(s) + iw_{oim}(s)$,

corresponding to the FF, SH and TH beams, respectively. To analyze the stability of these solutions with respect to small perturbations, we substitute $X_T(s,\xi) = X_0(s) + X(s)e^{\lambda\xi}$, where $X = (u_{re}, u_{im}, v_{re}, v_{im}, w_{re}, w_{im})^T$, into Eqs. (19), (20) and (21) and linearize the resulting equations obtaining a linear eigenvalue problem

$$LX = \lambda C, \tag{56}$$

Here $L$ is a $6 \times 6$ matrix having the following non-zero elements:

$$L_{11} = -\frac{r}{2}\frac{\partial^2}{\partial s^2} - \kappa_1 + d_{mn}v_{ore} + \frac{r_{11}}{2}(3u_{ore}^2 + u_{oim}^2) + \gamma_{12}(v_{ore}^2 + v_{oim}^2) + \gamma_{13}(w_{ore}^2 + w_{oim}^2) + 3\rho_{13}(u_{ore}w_{ore} + u_{oim}w_{oim})$$

,

$$L_{12} = L_{21} = d_{mn}v_{oim} + \gamma_{11}u_{oim}u_{ore} + 3\rho_{13}(u_{ore}w_{oim} - w_{ore}u_{oim}),$$

$$L_{13} = d_{mn}u_{ore} + d_{gh}w_{ore} + 2\gamma_{12}u_{ore}v_{ore} + 2\rho_{123}(v_{ore}w_{ore} + v_{oim}w_{oim}),$$

$$L_{14} = d_{mn}u_{oim} + d_{gh}w_{oim} + 2\gamma_{12}u_{ore}v_{oim} + 2\rho_{123}(v_{ore}w_{oim} - w_{ore}v_{oim}),$$

$$L_{15} = d_{gh}v_{ore} + 2\gamma_{13}u_{ore}w_{ore} + \frac{3}{2}\rho_{13}(u_{ore}^2 - u_{oim}^2) + \rho_{123}(v_{ore}^2 - v_{oim}^2),$$

$$L_{16} = d_{gh}v_{oim} + 2\gamma_{13}u_{ore}w_{oim} + 3\rho_{13}u_{ore}u_{oim} + 2\rho_{123}v_{ore}v_{oim},$$

$$L_{22} = -\frac{r}{2}\frac{\partial^2}{\partial s^2} - \kappa_1 - d_{mn}v_{ore} + \frac{r_{11}}{2}(u_{ore}^2 + 3u_{oim}^2) + \gamma_{12}(v_{ore}^2 + v_{oim}^2) + \gamma_{13}(w_{ore}^2 + w_{oim}^2) - 3\rho_{13}(u_{ore}w_{ore} + u_{oim}w_{oim})$$

,

$$L_{23} = -d_{mn}u_{oim} + d_{gh}w_{oim} + 2\gamma_{12}v_{ore}u_{oim} + 2\rho_{123}(w_{ore}v_{oim} - v_{ore}w_{oim}),$$

$$L_{24} = d_{mn}u_{ore} - d_{gh}w_{ore} + 2\gamma_{12}u_{oim}v_{oim} + 2\rho_{123}(v_{ore}w_{ore} + v_{oim}w_{oim}),$$

$$L_{25} = -d_{gh}v_{oim} + 2\gamma_{13}w_{ore}u_{oim} - 3\rho_{13}u_{ore}u_{oim} + 2\rho_{123}v_{ore}v_{oim},$$

$$L_{26} = d_{gh}v_{ore} + 2\gamma_{13}u_{oim}w_{oim} + \frac{3}{2}\rho_{13}(u_{ore}^2 - u_{oim}^2) + \rho_{123}(v_{oim}^2 - v_{ore}^2),$$

$$L_{31} = d_{-m-n}u_{ore} + d_{gh}w_{ore} + 2\gamma_{21}u_{ore}v_{ore} + 2\rho_{231}(v_{ore}w_{ore} + v_{oim}w_{oim}),$$

$$L_{32} = -d_{-m-n}u_{oim} + d_{gh}w_{oim} + 2\gamma_{12}v_{ore}u_{oim} + 2\rho_{231}(w_{ore}v_{oim} - v_{ore}w_{oim}),$$

$$L_{33} = -\frac{\alpha}{2}\frac{\partial^2}{\partial s^2} - \kappa_2 + \gamma_{21}(u_{ore}^2 + u_{oim}^2) + \gamma_{23}(w_{ore}^2 + w_{oim}^2) + 2\rho_{231}(u_{ore}w_{ore} - u_{oim}w_{oim}),$$

$$L_{34} = L_{43} = 2\rho_{231}(u_{ore}w_{oim} + w_{ore}u_{oim}),$$

$$L_{35} = d_{gh}u_{ore} + 2\gamma_{23}v_{ore}w_{ore} + 2\rho_{231}(u_{ore}v_{ore} + u_{oim}v_{oim}),$$

$$L_{36} = d_{gh}u_{oim} + 2\gamma_{23}v_{ore}w_{oim} + 2\rho_{231}(u_{ore}v_{oim} - v_{ore}u_{oim}),$$

$$L_{41} = d_{-m-n}u_{oim} + d_{gh}w_{oim} + 2\gamma_{21}u_{ore}v_{oim} + 2\rho_{231}(v_{ore}w_{oim} - w_{ore}v_{oim}),$$

$$L_{42} = d_{-m-n}u_{ore} - d_{gh}w_{ore} + 2\gamma_{21}u_{oim}v_{oim} + 2\rho_{231}(v_{ore}w_{ore} + v_{oim}w_{oim}),$$

$$L_{44} = -\frac{\alpha_2}{2}\frac{\partial^2}{\partial s^2} - \kappa_2 + \gamma_{21}(u_{ore}^2 + u_{oim}^2) + \gamma_{23}(w_{ore}^2 + w_{oim}^2) + 2\rho_{231}(u_{oim}w_{oim} - u_{ore}w_{ore}),$$

$$L_{45} = -d_{gh}u_{oim} + 2\gamma_{23}w_{ore}v_{oim} + 2\rho_{231}(v_{ore}u_{oim} - u_{ore}v_{oim}),$$

$$L_{46} = d_{gh}u_{ore} + 2\gamma_{23}v_{oim}w_{oim} + 2\rho_{231}(u_{ore}v_{ore} + u_{oim}v_{oim}),$$

$$L_{51} = d_{-g-h}v_{ore} + 2\gamma_{31}u_{ore}w_{ore} + \frac{3}{2}\rho_{31}(u_{ore}^2 - u_{oim}^2) + \rho_{312}(v_{ore}^2 - v_{oim}^2),$$

$$L_{52} = -d_{-g-h}v_{oim} + 2\gamma_{31}w_{ore}u_{oim} - 3\rho_{31}u_{ore}u_{oim} + 2\rho_{312}v_{ore}v_{oim},$$

$$L_{53} = d_{-g-h}u_{ore} + 2\gamma_{32}v_{ore}w_{ore} + 2\rho_{312}(u_{ore}v_{ore} + u_{oim}v_{oim}),$$

$$L_{54} = -d_{-g-h}u_{oim} + 2\gamma_{32}w_{ore}v_{oim} + 2\rho_{312}(v_{ore}u_{oim} - u_{ore}v_{oim}),$$

$$L_{55} = L_{66} = -\frac{\alpha_3}{2\chi}\frac{\partial^2}{\partial s^2} - \frac{\kappa_3}{\chi} + \gamma_{31}(u_{ore}^2 + u_{oim}^2) + \gamma_{32}(v_{ore}^2 + v_{oim}^2),$$

$$L_{56} = L_{65} = 0,$$

$$L_{61} = d_{-g-h}v_{oim} + 2\gamma_{31}u_{ore}w_{oim} + 3\rho_{31}u_{ore}u_{oim} + 2\rho_{312}v_{ore}v_{oim},$$

$$L_{62} = d_{-g-h}v_{ore} + 2\gamma_{31}u_{oim}w_{oim} + \frac{3}{2}\rho_{31}(u_{ore}^2 - u_{oim}^2) + \rho_{312}(v_{oim}^2 - v_{ore}^2),$$

$$L_{63} = d_{-g-h}u_{oim} + 2\gamma_{32}v_{ore}w_{oim} + 2\rho_{312}(u_{ore}v_{oim} - v_{ore}u_{oim}),$$

$$L_{64} = d_{-g-h}u_{ore} + 2\gamma_{32}v_{oim}w_{oim} + 2\rho_{312}(u_{ore}v_{ore} + u_{oim}v_{oim}), \tag{57}$$

and $\boldsymbol{C} = (u_{im}, -u_{re}, v_{im}, -v_{re}, w_{im}, -w_{re})^T$. For $\lambda = 0$ this eigenvalue problem has two spatially localized solutions $X_1 = \partial X_0/\partial s$ and $X_2 = (u_{oim}, -u_{ore}, 2v_{oim}, -2v_{ore}, 3w_{oim}, -3w_{ore})^T$ giving the neutrally stable modes of the linear eigenvalue problem. In order to find a threshold condition for the linear instability we introduce a small parameter $\varepsilon$, so that we can seek solutions of the above linear eigenvalue problem in the form of asymptotic series in the small parameter $\varepsilon$:

$$X = \sum_{j=0}^{\infty} \varepsilon^j X^{(j)}, \quad C = \sum_{j=0}^{\infty} \varepsilon^j C^{(j)}, \quad \lambda = \sum_{j=1}^{\infty} \varepsilon^j \lambda_j, \tag{58}$$

where $X^{(j)} = \left(u_{re}^{(j)}, u_{im}^{(j)}, v_{re}^{(j)}, v_{im}^{(j)}, w_{re}^{(j)}, w_{im}^{(j)}\right)^T$ and $C^{(j)} = \left(u_{im}^{(j)}, -u_{re}^{(j)}, v_{im}^{(j)}, -v_{re}^{(j)}, w_{im}^{(j)}, -w_{re}^{(j)}\right)^T$.

Above expansion is substituted into the linear problem of Eq. (56) to get, up to order $\varepsilon^2$:

$$O(1): LX^{(0)} = 0,$$

$$O(\varepsilon): LX^{(1)} = \lambda_1 C^{(0)},$$

$$O(\varepsilon^2): LX^{(2)} = \lambda_1 C^{(1)} + \lambda_2 C^{(0)}, \tag{59}$$

where $X_1^{(0)} = (\partial u_{ore}/\partial s, \partial u_{oim}/\partial s, \partial v_{ore}/\partial s, \partial v_{oim}/\partial s, \partial w_{ore}/\partial s, \partial w_{oim}/\partial s)^T$ and $X_2^{(0)} = (u_{oim}, -u_{ore}, 2v_{oim}, -2v_{ore}, 3w_{oim}, -3w_{ore})^T$ are two spatially localized solutions. For simplicity, we employ the latter solution in our following calculation.

We find the following explicit analytical solutions for the first-order corrections $X_2^{(1)} = \partial X_0/\partial \kappa_1$. The instability threshold condition emerges at the next, second-order in $\varepsilon$. Here we introduce the adjoint operator $L^+$ of $L$, which is also the transpose operator of $L$:

$$L^+ = L^T, \tag{60}$$

let $Y$ belong to the kernel space of the operator $L^+$,

$$L^+ Y = 0, \tag{61}$$

From the third of Eqs. (59), we obtain the solvability conditions.

$$Y^T C^{(0)} = 0, \tag{62}$$

$$Y^T C^{(1)} = 0, \tag{63}$$

where

$$Y_2 = (u_{oim}, -u_{ore}, 2v_{oim}, -2v_{ore}, 3w_{oim}, -w_{ore})^T,$$

$$C_2^{(0)} = (u_{ore}, u_{oim}, 2v_{ore}, 2v_{oim}, 3w_{ore}, 3w_{oim})^T,$$

$$C_2^{(1)} =$$
$$(\partial u_{oim}/\partial \kappa_1, -\partial u_{ore}/\partial \kappa_1, \partial v_{oim}/\partial \kappa_1, -\partial v_{ore}/\partial \kappa_1, (1/\chi) \partial w_{oim}/\partial \kappa_1, -(1/\chi) \partial w_{ore}/\partial \kappa_1)^T$$

.

From Eq. (63), we obtain the condition of marginal stability of the solutions

$$\frac{\partial I}{\partial \kappa_1} \equiv \frac{\partial}{\partial \kappa_1} \int \left( u_{ore}^2 + u_{oim}^2 + 2v_{ore}^2 + 2v_{oim}^2 + \frac{3}{\chi} w_{ore}^2 + \frac{3}{\chi} w_{oim}^2 \right) ds = 0, \qquad (64)$$

which is the Vakhitov-Kolokolov (VK) criterion extensively applied to various classes of solitary waves [27,34].

## V. CONCLUSION

We have analyzed the (1+1)-dimensional multicolor solitary waves due to cascading quadratic nonlinear response in general one-dimensional quasiperiodic optical superlattices. We have shown the dynamic equations describing multicolor quasiperiodic solitons in QPM quadratic media include quasiperiodicity-induced Kerr effects, such as self- and cross-phase modulation, THG and four-wave mixing. We have numerically predicted spatial self-trapping of quasiperiodic waves and the formation of quasiperiodic envelope solitons which propagate for more than twenty diffraction lengths without any significant loss of power. The stability of this multicolor soliton has been demonstrated in the framework of the average dynamical equations by means of a Lyapunov analysis based on the energy integral of the wave coupling equations. We have investigated the dynamics of the multicolor solitons using a virial identity, which predicts a stable propagation of the mutually trapped solitons. We have established the analytic stability criterion for the multicolor solitons by applying a multiscale asymptotic method.

**Figure 1**

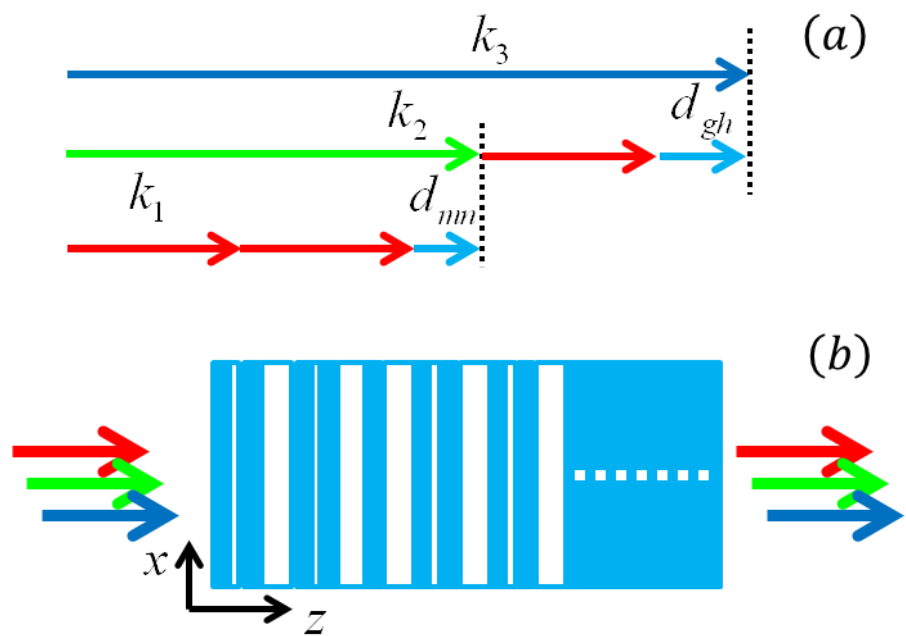

Fig1. (a) Schematic Presentation for two phase-matching conditions involved in the Second Harmonic Generation (SHG) and the Sum-Frequence Mixing (SFM) processes. (b) The general quasiperiodic quasi-phase-matched grating used for the formation of multicolor solitons.

**Figure 2**

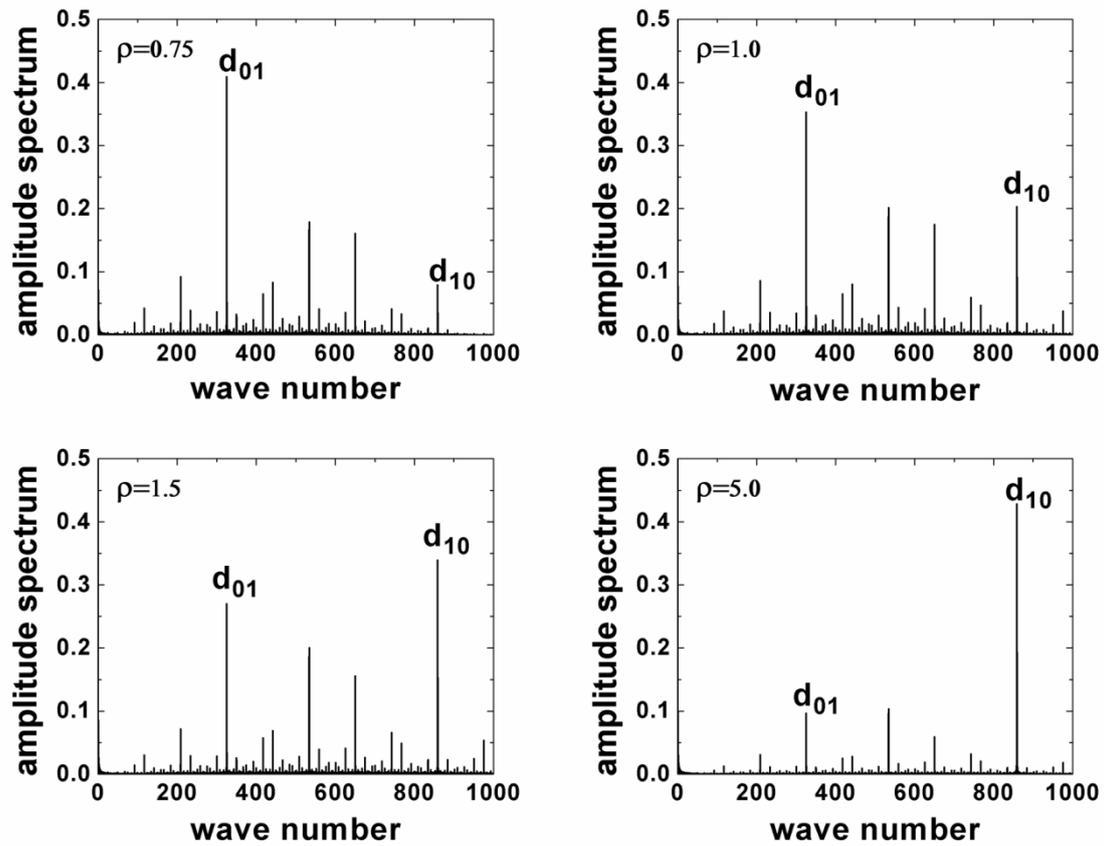

FIG. 2. Numerically calculated amplitude spectrum of $d(z)$ for $C_b = 0.8$ and (a) $\rho = 0.75$, (b) $\rho = 1$, (c) $\rho = 1.5$, (d) $\rho = 5$. $d_{01}$ and $d_{10}$ are the Fourier transforms of two reciprocal vectors quasi-phase-matching for the SHG and SFG processes, respectively.

**Figure 3**

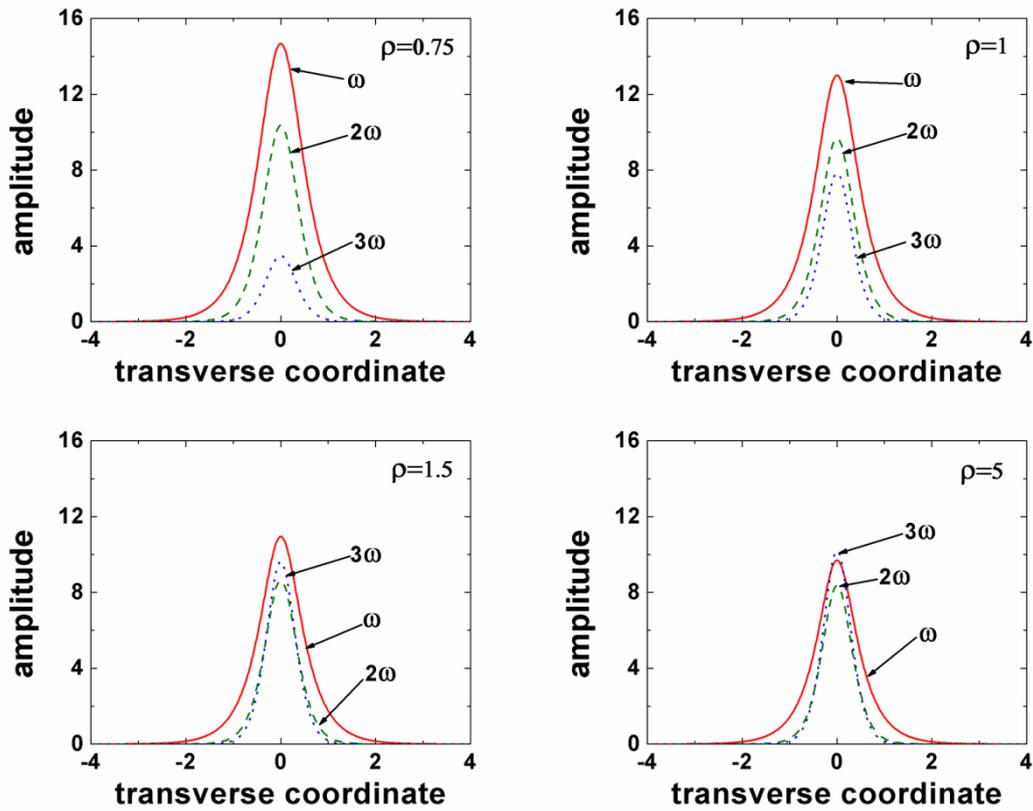

FIG. 3. Amplitude profile of quasiperiodic QPM multicolor solitons calculated for $C_b = 0.8$ and (a) $\rho = 0.75$, (b) $\rho = 1$, (c) $\rho = 1.5$, (d) $\rho = 5$. The curves plot the fundamental frequency $\omega$ (solid curves), the second harmonic $2\omega$ (dashed curves), and the third harmonic $3\omega$ (dotted curves) fields along the transverse coordinate *s*.

**Figure 4**

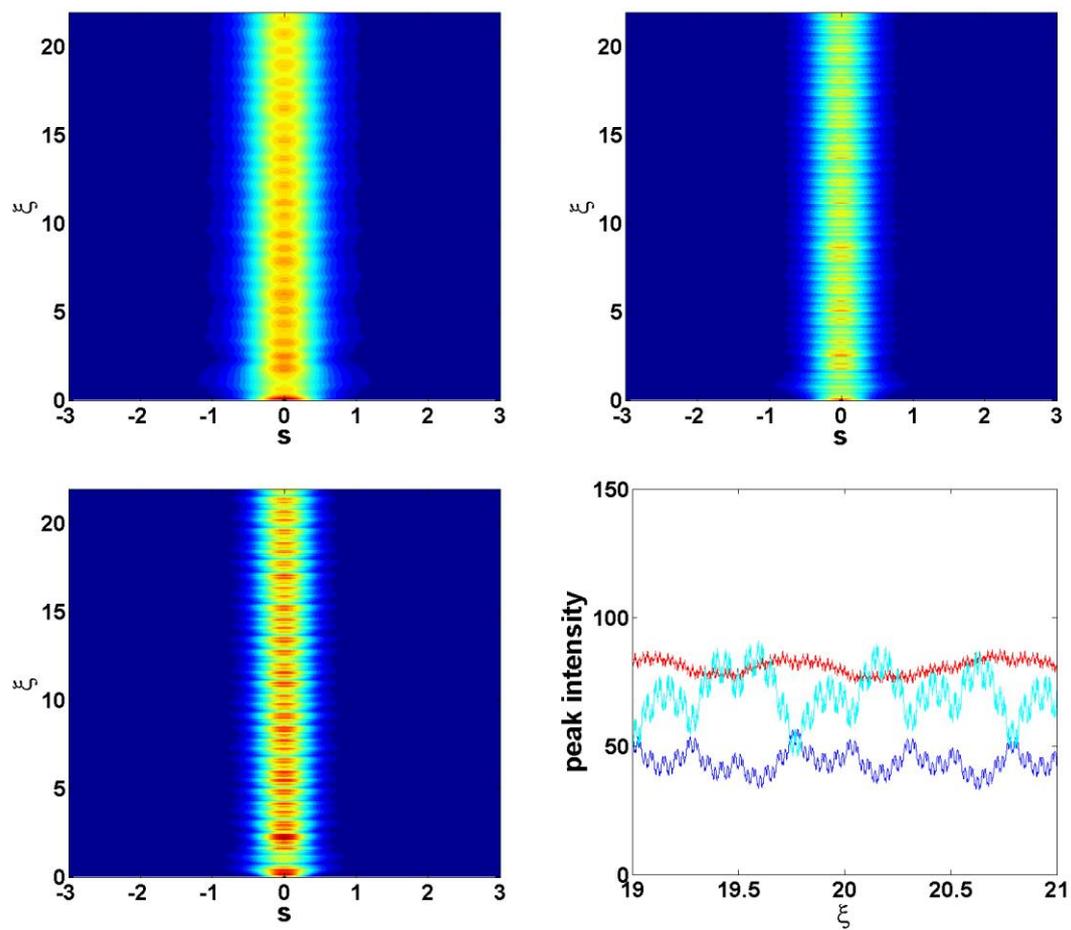

FIG. 4. Evolution of multicolor soliton intensities upon propagation in a generalized quasiperiodic QPM waveguide grating for (a) the fundamental frequency $\omega$, (b) the second harmonic $2\omega$, and (c) the third harmonic $3\omega$. (d) Peak intensity oscillations of the quasiperiodic multicolor soliton.

**Figure 5**

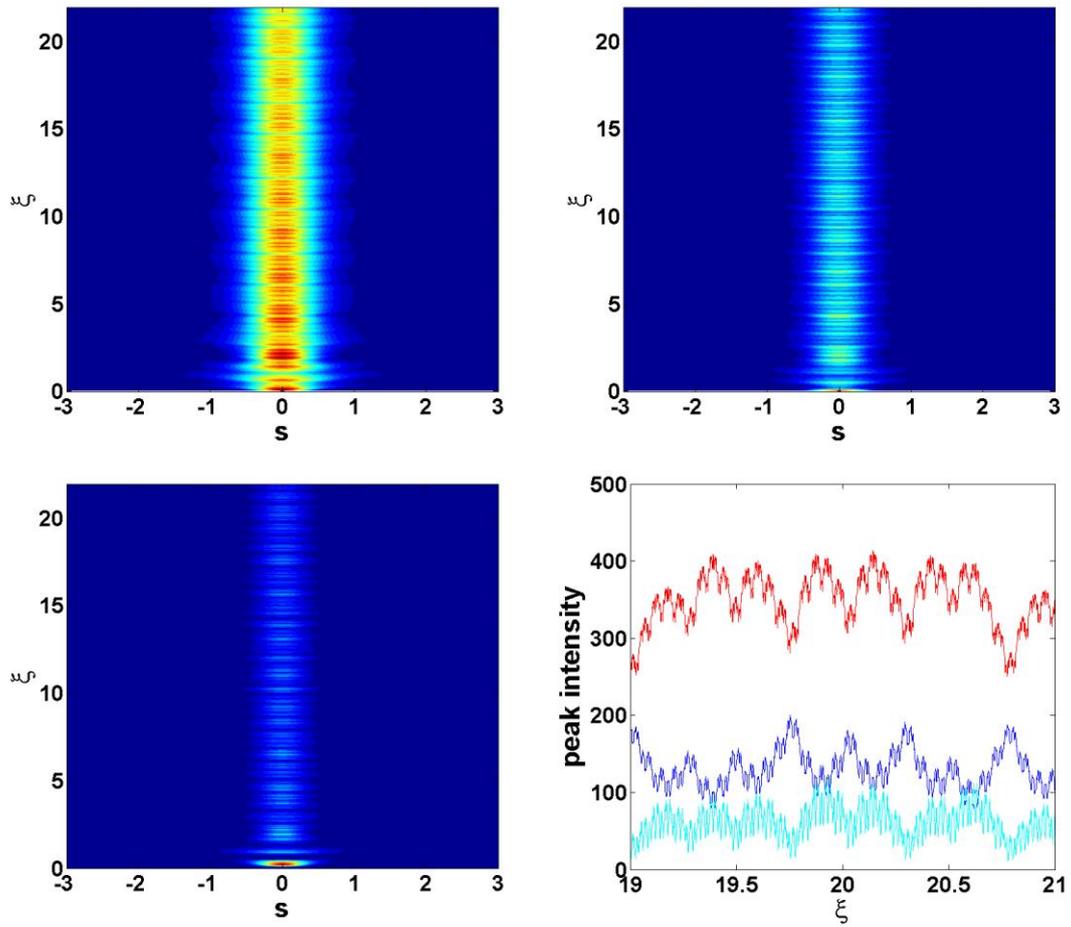

FIG. 5. Excitation of multicolor soliton upon propagation in a generalized quasiperiodic QPM waveguide grating for (a) the fundamental frequency $\omega$, (b) the second harmonic $2\omega$, and (c) the third harmonic $3\omega$. (d) Peak intensity oscillations of the quasiperiodic multicolor soliton.